\documentclass[aps,prl,superscriptaddress,amssymb,reprint]{revtex4-1}
\pdfoutput=1
\usepackage{amsmath, amsfonts}    
\usepackage{graphicx}   
\usepackage{verbatim}   
\usepackage{color}      
\usepackage{subfigure}  
\usepackage{hyperref}   
\usepackage{natbib}
\usepackage[USenglish]{babel}
\usepackage{placeins} 

\usepackage{epsfig}
\newcommand{\changefont}[3]{\fontfamily{#1} \fontseries{#2} \fontshape{#3} \selectfont}

\begin{document}

\title{High-fidelity projective readout of a solid-state spin quantum register}

\author{Lucio Robledo}
\email{These authors contributed equally to this work}
\affiliation{Kavli Institute of Nanoscience Delft, Delft University of Technology, P.O. Box 5046, 2600 GA Delft, The Netherlands}
\author{Lilian Childress}
\email{These authors contributed equally to this work}
\affiliation{Department of Physics and Astronomy, Bates College, 44 Campus Ave., Lewiston, ME 04240, USA}
\author{Hannes Bernien}
\email{These authors contributed equally to this work}
\affiliation{Kavli Institute of Nanoscience Delft, Delft University of Technology, P.O. Box 5046, 2600 GA Delft, The Netherlands}
\author{Bas Hensen}
\affiliation{Kavli Institute of Nanoscience Delft, Delft University of Technology, P.O. Box 5046, 2600 GA Delft, The Netherlands}
\author{Paul F. A. Alkemade}
\affiliation{Kavli Institute of Nanoscience Delft, Delft University of Technology, P.O. Box 5046, 2600 GA Delft, The Netherlands}
\author{Ronald Hanson}
\email{r.hanson@tudelft.nl}
\affiliation{Kavli Institute of Nanoscience Delft, Delft University of Technology, P.O. Box 5046, 2600 GA Delft, The Netherlands}

\begin{abstract}
Initialization and readout of coupled quantum systems are essential ingredients for the implementation of 
quantum algorithms \cite{Nielsen2000, Raussendorf2003}. If the state of a multi-qubit register can be read out in a single 
shot, this enables further key resources such as quantum error correction 
and deterministic quantum teleportation \cite{Nielsen2000}
, as well as direct investigation of quantum correlations (entanglement).  While spins in solids are attractive candidates for scalable quantum information processing, thus far single-shot detection has only been achieved for isolated qubits~\cite{Elzerman2004, 
Vamivakas2010,Neumann2010,Morello2010}.   Here, we demonstrate preparation and measurement of a multi-spin quantum register by implementing resonant optical excitation techniques originally developed in atomic physics.  We achieve high-fidelity 
readout of the electronic spin associated with a single nitrogen-vacancy (NV) centre in diamond at low temperature, and exploit this readout to project  up to three nearby nuclear spin qubits onto a well-defined state~\cite{Giedke2006}. Conversely, we can distinguish the state of the nuclear spins in a single shot by mapping it onto and subsequently measuring the electronic spin~\cite{Jiang2009, Neumann2010}.  Finally, we show compatibility with qubit control by demonstrating initialization, coherent manipulation, and single-shot readout  in a single experiment  on a two-qubit register, using techniques suitable for extension to larger registers.  These results pave the way for the first test of Bell's inequalities on solid-state spins and the implementation of measurement-based quantum information protocols.
\end{abstract}

\maketitle

\begin{figure}
  \includegraphics{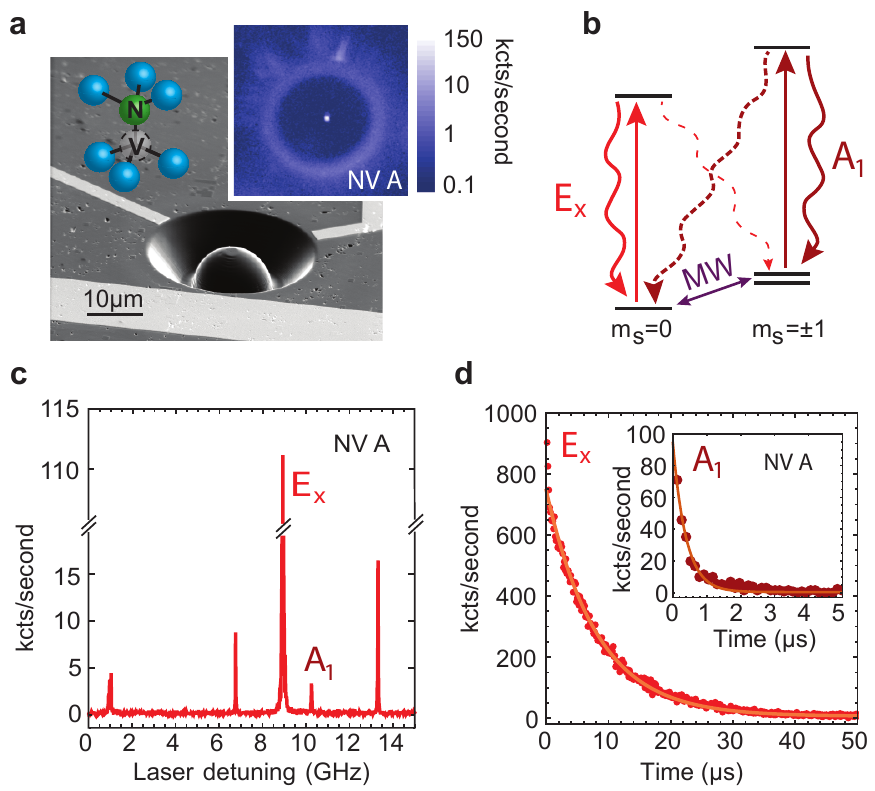}
  \caption{\label{fig:fig1}$|${\changefont{phv}{b}{n}Resonant excitation and electronic spin preparation of an NV centre.}{\changefont{phv}{b}{n}a},~Scanning electron microscope image of a solid immersion lens (SIL) representative of those used in the experiments (for details see Supplementary Information). The overlayed sketch shows the substitutional nitrogen and adjacent vacancy that form the NV centre. Inset: Scanning confocal microscope image of NV A (log. colour scale).{\changefont{phv}{b}{n}b},~Energy levels used to prepare and read out the NV centre's electronic spin ($S=1$ in the ground and optically excited states); transitions are labelled according to the symmetry of their excited state. 
Dashed lines indicate spin-non-conserving decay paths.{\changefont{phv}{b}{n}c}, Photoluminescence excitation spectrum of NV~A, frequency is given relative to 470.443\,THz.{\changefont{phv}{b}{n}d},~Fluorescence time trace of NV A, initially prepared in $m_s=0$ ($E_x$ excitation, $P = 4.8$\,nW) and $m_s=\pm1$ ($A_1$ excitation, $P = 7.4$\,nW, inset), $P_{sat} \approx 6$\,nW. Spin flips in the excitation cycle lead to nearly exponential decay of fluorescence, with a fitted spin-flip time of $1/\gamma_0=8.1\pm 0.1\,\mu$s ($0.39 \pm 0.01\,\mu$s) for $E_x$ ($A_1$), and an initial intensity of $740\pm5$ ($95\pm2$) kcounts per second, giving a lower limit to the $m_s=0$ and $m_s=\pm1$ preparation fidelity of 99.7\,$\pm$\,0.1\,\% and 99.2\,$\pm$\,0.1\,\%, respectively.  The low initial counts on $A_1$ are associated with a fast intersystem crossing to metastable singlet states (see Supplementary Information).}
\end{figure}

The electronic spin of the NV centre in diamond constitutes an exceptional solid state system for investigating quantum phenomena, combining excellent spin coherence~\cite{Balasubramanian2009,Naydenov2011,deLange2010,Ryan2010} 
with a robust optical interface\cite{Togan2010,Batalov2008,Buckley2010,Robledo2010}.  Furthermore, the host nitrogen nuclear spin (typically $^{14}$N, $I=1$) and proximal isotopic impurity $^{13}$C nuclei ($I=\frac{1}{2}$) exhibit hyperfine interactions with the NV electronic spin ($S=1$), enabling development of few-spin quantum registers that have been envisioned as building blocks for quantum repeaters~\cite{Childress2006a}, cluster state computation~\cite{Barrett2005}, and distributed quantum computing~\cite{Jiang2007}. All of these applications require high-fidelity preparation, manipulation, and measurement of multiple spins. 
While there have been significant advances in coherent control over few-spin systems in diamond~\cite{
Dutt2007, Neumann2008}, 
no method exists for simultaneous preparation~\cite{Fuchs2008, Jacques2009} and single-shot readout~\cite{Neumann2010} of multi-spin registers, impeding progress towards multi-qubit protocols. Here, we remove this roadblock by exploiting 
 resonant excitation techniques, as pioneered in atomic physics~\cite{Happer1972, Blatt1988}, in micro-structured diamond devices that enable high photon collection efficiency (Fig.~\ref{fig:fig1}a). These new methods enable us to initialize multiple nuclear spin qubits, and to perform single-shot readout of a few-qubit register, clearing the way towards implementation of quantum algorithms with solid-state spins.

Our preparation and readout techniques rely on resonant excitation of spin-selective optical transitions of the NV centre, which can be spectrally resolved at low temperatures~\cite{Tamarat2008}. We use the $E_x$ and $A_1$ transitions in our experiments (see Fig.~\ref{fig:fig1}b): $A_1$ connects the ground states with $m_s = \pm1$ spin projection to an excited state with primarily $m_s = \pm 1$ character, whereas $E_x$ connects states with $m_s = 0$. A typical spectrum from NV A, one of the two NVs we study, is shown in Fig.~\ref{fig:fig1}c (see also Supplementary Information). 
Under resonant excitation of a single transition, the fluorescence decays with time owing to a slight spin mixing within the excited states that induces shelving into the other spin state (Fig.~\ref{fig:fig1}d).  This optical pumping mechanism enables high-fidelity spin state initialization\cite{Happer1972,Atature2006}: from the data in Fig.~\ref{fig:fig1}d we estimate 
a preparation error into the $m_s=0$ ground state of 0.3\,$\pm$\,0.1\,\%, which is a drastic reduction of the 11\,$\pm$\,3\,\% preparation error observed with conventional off-resonant initialization (see Supplementary Information). 

Spin-dependent resonant excitation also allows single-shot electronic spin readout: the presence or absence of fluorescence under $E_x$ excitation reveals the spin state.  By working with low-strain NVs at low temperature ($T=8.6$\,K), we suppress spin-mixing~\cite{Manson2006,Tamarat2008} and phonon-induced transitions~\cite{Fu2009} within the excited states, extending the spin relaxation time under $E_x$ excitation to several microseconds. Together with a high collection efficiency due to the use of solid immersion lenses~\cite{Hadden2010} fabricated around pre-selected, low-strain NVs, and efficient rejection of resonant excitation from the measured phonon-sideband emission, this highly spin-preserving transition allows the detection of several photons before the spin flips.  

\begin{figure*}
	\includegraphics{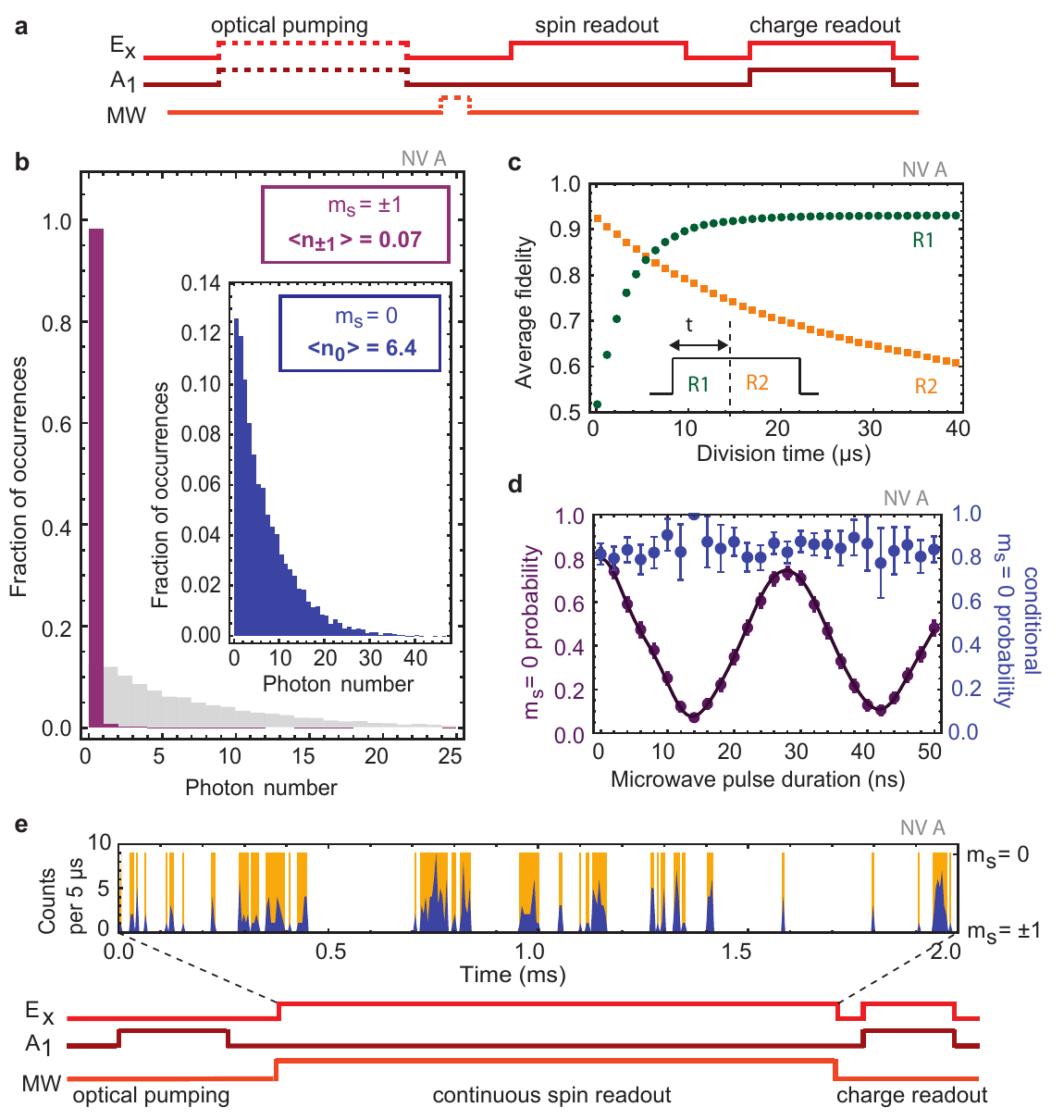}
  \caption{\label{fig:fig2}$|${\changefont{phv}{b}{n}Projective single-shot readout of the NV electronic spin.}{\changefont{phv}{b}{n}a},~Pulse sequence used for electronic spin readout: after charge initialization (532\,nm, not shown) the electron is pumped into $m_s=0$ ($A_1$, dark red) or $m_s=\pm1$ ($E_x$, bright red), followed by optional microwave (MW) spin manipulation and the spin readout pulse resonant with $E_x$.  Conditioning on simultaneous resonance during the final charge and detuning sensing stage eliminates effects of local electric field fluctuations or ionization (see Supplementary Information).   {\changefont{phv}{b}{n}b},~Statistics of photon counts detected during a $t_{ro}=100\,\mu$s electronic spin readout after initialization into $m_s=\pm1$ (red) and $m_s=0$ (superimposed light gray and inset), obtained from 10000 measurement repetitions.{\changefont{phv}{b}{n}c},~When the $100\,\mu$s readout pulse is divided into two readout segments R1 and R2 with variable division point, the fidelity of two consecutive segments reaches $83.4\pm0.5\%$ for an optimal division time of $5.5\,\mu$s; the probability of identical sequential outcomes is $82.0\pm 0.7\%$. 2 SE error bars (n = 10000) are smaller than the symbols. {\changefont{phv}{b}{n}d},~Electronic spin Rabi oscillations between $m_s=0$ and $m_s=-1$ at $B_z \approx 15$\,G (purple): each data point is obtained from 1000 single-shot readout repetitions. The fit, which includes the detailed hyperfine level structure, yields a visibility of $78\pm 8\%$ , where a maximum of $84\%$ can be expected. Blue data points show the measurement outcome after projection into $m_s=0$ by selecting only readout events with photons detected within the first 400\,ns (see Supplementary Information). All errors and error bars are 2 SE. {\changefont{phv}{b}{n}e},~Quantum jumps in the fluorescence time-trace during continuous spin readout. Durations of dark periods depend on the MW Rabi frequency (see Supplementary Information). Blue data indicates the counts per readout bin of $5\,\mu s$, and the deduced spin state is shown in orange.}
\end{figure*}

We demonstrate single-shot readout by initializing the electronic spin into $m_s=0$ or $m_s=\pm1$, followed by resonant excitation on the $E_x$ readout transition for $t_{ro}=100$\,$\mu$s (Fig.~\ref{fig:fig2}a). The resulting histograms of the number of detected photons are given in Fig.~\ref{fig:fig2}b. As expected, for $m_s=\pm1$ we observe negligible excitation, with a 98.3\,\% probability to not measure any photon during the probe interval. In stark contrast, after initialization into $m_s=0$ we detect on average $\langle n_0 \rangle = 6.4$ photons per shot. 
We assign the state $m_s=0$ to detection of one or more photons, and $m_s=\pm1$ otherwise. After truncating our integration window to the optimal duration of $40\,\mu$s, we find an average fidelity $F_{\mathrm{avg}}=\frac{1}{2}(F_{m_s=0}+F_{m_s=\pm1}) =93.2\pm0.5\%$; here $F_{m_s}$ is the probability to obtain the measurement outcome $m_s$ after optical pumping into $m_s$. To verify that these measurement outcomes indeed correspond to the electronic spin states, we utilise single-shot readout to observe spin Rabi 
oscillations and microwave-induced quantum jumps~\cite{Blatt1988} (see Fig.~\ref{fig:fig2}d,e). 

While the full readout optically pumps the spin, shorter readout durations can be non-destructive, albeit at lower fidelity.   
By optimizing integration windows, we obtain a fidelity of $83.4\pm0.5$\% for each of two successive readout segments (Fig.~\ref{fig:fig2}c). Correlations between measurement outcomes indicate that the readout is projective. Following preparation of a superposition of spin states, we condition on detection of at least one photon (i.e., measurement outcome $m_s = 0$) during a first short readout pulse, and probe the resulting spin state with a second readout (Fig.~\ref{fig:fig2}d, blue data points). Regardless of the initial spin state, we observe a constant high probability to obtain $m_s = 0$ in the second readout. This shows that the readout method is strongly projective and well suited for application in measurement-based quantum protocols.

\begin{figure*}
  \includegraphics{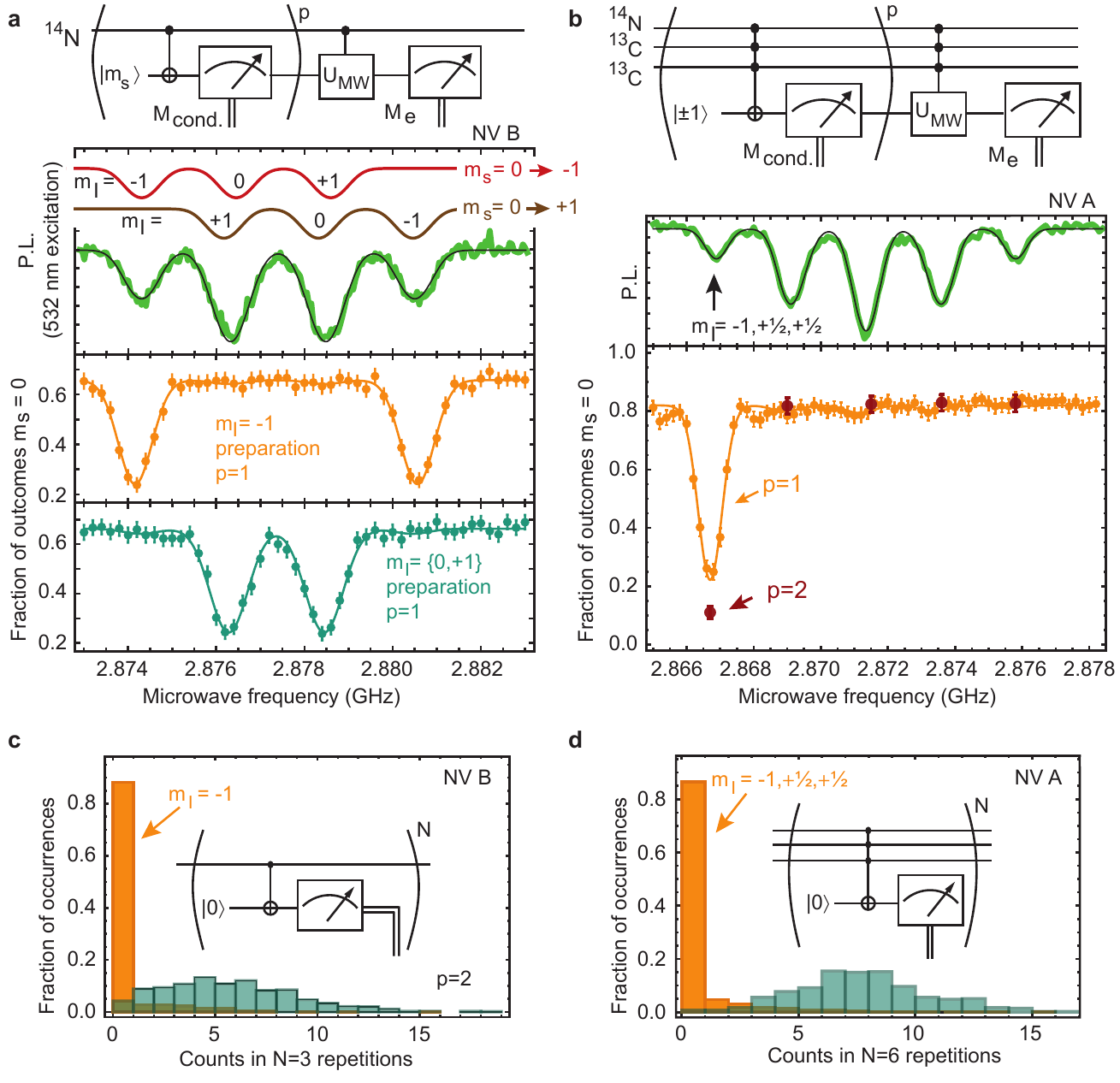}
  \caption{\label{fig:fig3}$|${\changefont{phv}{b}{n}Nuclear spin preparation and readout.}{\changefont{phv}{b}{n}a},~Measurement-based preparation of a single $^{14}$N nuclear spin. In the ambient Earth magnetic field of $\approx 0.5$\,G, without nuclear spin polarization, we observe four resonances in the hyperfine spectrum (green trace) for NV B;  the outer two correspond to the nuclear spin state $m_I=-1$ and the central two are combinations of $m_I=\{0,+1\}$.
Circuit diagram: To initialize the nuclear spin we entangle it with the electronic spin and then read out the latter. Data for $m_I=-1$ preparation is shown in orange, preparation in $m_I=\{0,+1\}$  in cyan (below). Fits to Gaussian spectra show an amplitude ratio of $96\pm4$\% in the desired nuclear spin state.  {\changefont{phv}{b}{n}b},~Measurement-based preparation of a three nuclear spin register. Using a similar sequence (circuit diagram) we prepare a well-defined state for all three nuclear spins.  A portion of the un-initialized hyperfine spectrum (green) contains 12 partially superposed lines, of which we prepare the single line corresponding to $m_I=(-1, \frac{1}{2}, \frac{1}{2})$ (orange, below).  Gaussian fits constrained to known hyperfine splittings yield an amplitude ratio of $88\pm10$\%.  The observed visibility can be improved by performing two preparation steps and electronic spin repumping ($p = 2$, five red data points), increasing the contrast to 82\% of the expected visibility from known readout fidelity (see Supplementary Information).  {\changefont{phv}{b}{n}c},~Single-shot measurement of the $^{14}$N nuclear spin, preceded by two preparation steps ($p=2$).  Readout (3 repetitions) conditioned on successful preparation distinguishes $m_I=-1$ (orange, threshold $<$ 1 count) from $m_I=\{0,+1\}$ (cyan) with an average fidelity of $92\pm2$\,\% (see Supplementary Information). {\changefont{phv}{b}{n}d},~Multiple nuclear spin readout. Using a similar sequence (inset) we distinguish one of the 12 hyperfine states associated with NV~A. To prepare nuclear spin states, we perform the readout procedure $7$ times and keep only data with zero total counts  (identified as $(-1, \frac{1}{2}, \frac{1}{2}))$ or $\geq 2$ counts per initialization step (other states).  Subsequent readout with 6 repetitions $((-1, \frac{1}{2}, \frac{1}{2})$  threshold $<$ 3 total counts) achieves a 96.7$\pm$ 0.8\% average fidelity for preparation and detection of the nuclear spin configuration. Uncertainties and error bars are 2 SE (n = 1000 for{\changefont{phv}{b}{n}a,b}, $p=1$; selected from 10000 measurement runs for {\changefont{phv}{b}{n}b}, $p=2$).
}
\end{figure*}

We exploit projective readout of the electron spin in combination with quantum gate operations for initialization and readout of a nuclear spin few-qubit register. We first demonstrate the concept of measurement-based preparation on a single nuclear qubit. The electronic spin resonance (ESR) spectrum for NV B (Fig.~\ref{fig:fig3}a, green trace) reveals the coupling to the host $I = 1 ~^{14}$N nuclear spin: two partially overlapping sets of three hyperfine lines correspond to the $m_s=0\leftrightarrow-1$ and $m_s=0\leftrightarrow+1$ electronic spin transitions, Zeeman-split by $\sim$\,2\,MHz in the earth's magnetic field. 
The outermost transitions are associated with a specific nuclear spin state $m_I$, e.g. $(m_s,m_I)=(0,-1) \leftrightarrow (-1,-1)$ at 2.874\,GHz.  
Our initialization procedure works as follows (Fig.~\ref{fig:fig3}a, circuit diagram): first, we prepare the electronic spin in $m_s=\pm1$. We then perform a nuclear-spin-controlled NOT gate on the electronic spin by applying a $\pi$-pulse at 2.874\,GHz; this operation rotates the electronic spin into $m_s=0$ only when $m_I=-1$. Finally, we read out the electronic spin for 400\,ns. If one or more photons are detected during this interval, the two-spin system is projected into ($m_s,m_I)=(0,-1)$. Alternatively, if we run the same protocol with initial electronic spin state $m_s=0$, we prepare the nuclear spin into $m_I=\{0,+1\}$.

The efficiency of the nuclear spin initialization is evidenced by its dramatic effect on the ESR spectrum (Fig.~\ref{fig:fig3}a). Whereas before preparation the depths of the different hyperfine lines indicate an equal mixture of the nuclear spin states (green trace), after preparation only the hyperfine lines corresponding to the prepared states are visible ($m_I=-1$ for cyan trace and $m_I=\{0,+1\}$ for the orange trace).

The same nuclear spin initialization scheme can be applied to multi-qubit registers. Figure~\ref{fig:fig3}b displays the ESR spectrum of NV A (green trace), whose electronic spin is coupled to both the host $^{14}$N nuclear spin and two nearby $^{13}$C nuclei (see Supplementary Information). The lowest-frequency line corresponds to a single state of the three nuclear spins. A $\pi$-pulse on this transition therefore implements a triple-controlled-NOT gate on the electronic spin (circuit diagram in Fig.~\ref{fig:fig3}b), enabling the initialization of all three nuclear spins (Fig.~\ref{fig:fig3}b, orange trace). The initialization can be further improved by repeating the preparation step (Fig.~\ref{fig:fig3}b, red data points). 

The nuclear qubits can be read out in a single shot by applying a nuclear-controlled-NOT on the electronic spin and subsequently reading out the electronic spin (see Fig.~\ref{fig:fig3}c,d insets). Because the electronic spin measurement only has weak back-action on the nuclear spin, we can repeat the process to obtain higher readout fidelity~\cite{Jiang2009,Neumann2010}. Figure~\ref{fig:fig3}c compares the resulting photon statistics for NV~B after initialization into the single nuclear spin state $m_I=-1$ to those obtained for $m_I=\{0,+1\}$, indicating an average readout fidelity of $92\pm2$\,\%. This number is a lower bound to the true readout fidelity, as it includes errors in the state preparation. 

A straightforward extension of this scheme can be used to read out the complete state of a multi-nuclear-spin register. Using a multiply-controlled-NOT gate in the readout sequence, we can measure in a single shot whether the register is in a particular configuration. We demonstrate this procedure on NV A, where we identify the 3-nuclear-spin state $m_I = (-1, \frac{1}{2}, \frac{1}{2})$ (see Fig. 3d and Supplementary Information). The other possible configurations can be probed by sequential application of this readout scheme on different spectrally-resolved hyperfine transitions, or, alternatively, by systematically flipping the nuclear spin qubits and repeating the readout on the same hyperfine transition.

Electron-nuclear flip-flop processes in the optically excited state, which reduce the nuclear spin readout fidelity, pose a major hurdle for scaling the readout to more qubits. Critically, resonant readout allows selection of which states undergo optical excitation. By starting with the electronic spin in $m_s=\pm1$, optical excitation will only occur when the register is in the state being probed; therefore, no optically-induced nuclear spin flips will occur during measurement of any of the other states. Thus, in contrast to schemes depending on off-resonant excitation where each additional readout step degrades the fidelity, resonant excitation enables scaling of high-fidelity readout to larger registers. 

\begin{figure*}
  \includegraphics{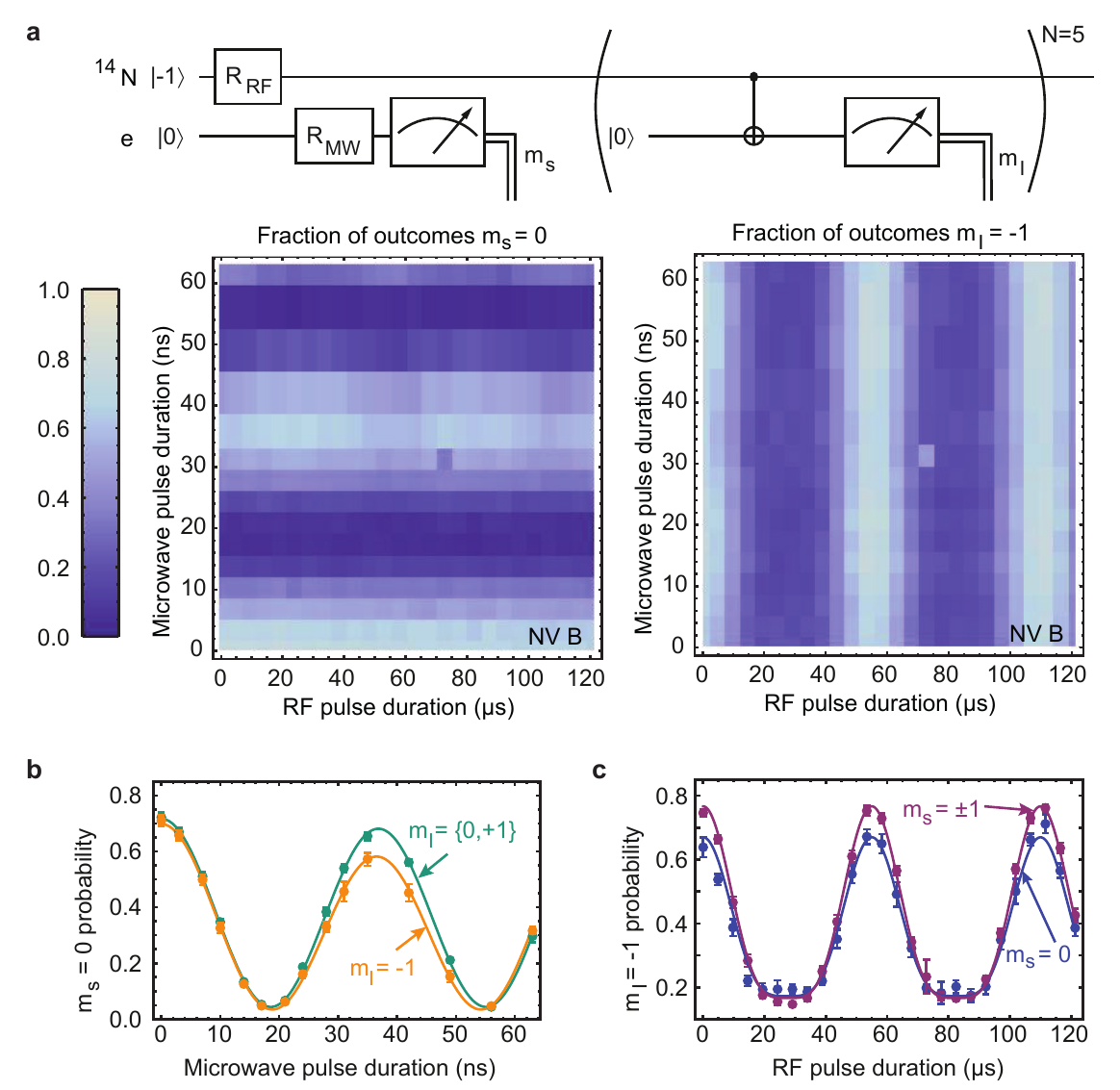}
  \caption{\label{fig:fig4}$|${\changefont{phv}{b}{n}Initialization, manipulation and readout of a two-qubit register.}{\changefont{phv}{b}{n}a},~After initialization of NV B into $(m_s,m_I)=(0,-1)$, we use radio-frequency excitation (RF, 4.9464\,MHz) to drive the nuclear spin and then microwaves (MW, 2.8774\,GHz) to drive the electronic spin. The electronic spin state is subsequently measured for $15\,\mu$s, followed by 5 readout steps (each $10\,\mu$s) of the $^{14}$N nuclear spin state.{\changefont{phv}{b}{n}b},~Probabilities to observe $m_s=0$ conditional on the measured nuclear spin state and averaged over all RF pulse durations, as a  function of MW pulse duration (see Supplementary Information).{\changefont{phv}{b}{n}c},~Probabilities for observing $m_I=-1$ conditional on the observed electronic spin state and averaged over all MW pulse durations, as a function of RF pulse duration (see Supplementary Information). 
All error bars and uncertainties are 2 SE; data based on 1000 measurements per pixel.
 }
\end{figure*}

Finally, we demonstrate the compatibility of all the different techniques by implementing them in a single experiment: we initialize, coherently manipulate, and then read out a two-qubit register consisting of the electronic spin and $^{14}$N nuclear spin of NV~B. After initialization 
in $(m_s,m_I)=(0,-1)$, we rotate the nuclear spin using a radiofrequency (RF) pulse and subsequently rotate the electronic spin with a microwave (MW) pulse. We then read out the electronic spin, followed by readout of the $^{14}$N nuclear spin state (Fig.~\ref{fig:fig4}a, circuit diagram). The left panel in Fig.~\ref{fig:fig4}a displays the readout results for the electronic spin qubit, showing the Rabi oscillations as a function of MW pulse length (vertical axis). In contrast, the readout results for the nuclear spin qubit (right panel of Fig.~\ref{fig:fig4}a) exhibit Rabi oscillations as a function of the RF pulse length (horizontal axis). 

To quantify crosstalk, we closely examine correlations between the two measurement outcomes. We observe that the contrast in the electronic spin Rabi oscillations depends on the outcome of the nuclear readout (Fig.~\ref{fig:fig4}b), but this discrepancy can be fully accounted for by the finite MW power 
used in this experiment (see Supplementary Information). The observed correlations thus arise from imperfect manipulation rather than measurement crosstalk. 
In the other direction, however, true measurement crosstalk appears: nuclear Rabi oscillation amplitudes decrease 
when the electronic spin is measured to be in $m_s=0$ (Fig.~\ref{fig:fig4}c) because optical excitation during electronic spin readout (which only succeeds for $m_s=0$) induces nuclear spin relaxation (see Supplementary Information for details).  
This effect can be mitigated by improving the collection efficiency (thus reducing the readout duration), e.g. by integrating the NV centre in an optical cavity. Also, application of moderate magnetic fields can cut the optically-induced nuclear spin relaxation rate by orders of magnitude~\cite{Neumann2010}. 

Our results have implications for a broad range of spin-based applications.  Single-shot electron spin readout can drastically improve NV-based sensors by enabling fast, quantum projection limited detection, creating opportunities in low-temperature magnetometry~\cite{Degen2008, Taylor2008}.  Extension of nuclear spin preparation techniques to remote nuclei in the spin bath may permit line-narrowing for enhanced sensitivity to d.c. magnetic fields.   Furthermore, preparation, manipulation, and single-shot readout of two spins open the door to exploration of two-particle quantum correlations, such as Bell's inequalities, and elementary quantum information processing protocols. Importantly, the techniques we describe are 
extendable to larger spin registers, and can be combined with precise spin qubit control and dynamical decoupling for coherence protection~\cite{deLange2010,Ryan2010,Naydenov2011}. The preparation and readout fidelities reported here are sufficient for demonstrating measurement-based entanglement generation and quantum teleportation of spin qubits, and exploring elementary quantum error correction schemes~\cite{Nielsen2000}. Ultimately, integration of multi-spin registers with quantum optical channels via spin-photon entanglement~\cite{Togan2010} may enable their application as few-qubit nodes in long-distance quantum communication protocols or distributed quantum information processing networks.

\FloatBarrier

\section{Methods}
All data were obtained by detecting photons emitted into the phonon sideband (650 - 750\,nm). For photoluminescence excitation spectroscopy, 5.5\,nW of red excitation is applied while microwaves at 2.878\,GHz coupled through an on-chip stripline drive the electronic spin transitions to prevent optical pumping. Scans are recorded in a single laser-frequency sweep at \mbox{$\sim200$\,MHz/s,} and are preceded by a 10\,$\mu$s pulse of 532\,nm excitation (50\,$\mu$W).  The green light is necessary to reset the negative charge state of the NV center, which can be photo-ionized by continuous resonant excitation.  For all other experiments, 532-nm-induced spectral diffusion must also be controlled: to ensure that the NV is on resonance with the red excitation lasers, we condition our data on strong fluorescence upon simultaneous $E_x$ and $A_1$ excitation following the experimental sequence (details in Supplementary Information). All errors and error bars are two standard error statistical uncertainty in the mean (95\% confidence interval).

\section{Acknowledgements} L.R. acknowledges support by a Marie Curie Intra European Fellowship within the 7$^{th}$ European Community Framework Programme. L.R., H.B., B.H. and R.H. gratefully acknowledge support from the Dutch Organization for Fundamental Research on Matter (FOM) and the European Commission (SOLID). L.C. acknowledges support from Research Corporation.

\section{Author Contributions} L.R., L.C. and H.B. conducted the experiments. L.R., L.C., H.B., B.H. and R.H. analysed the data. H.B. and P.F.A.A. fabricated the devices. L.R., L.C. and R.H wrote the paper. All authors commented on the manuscript. L.R., L.C. and H.B. contributed equally to this work.

\end{document}